\documentclass[conference,letterpaper]{IEEEtran}
\IEEEoverridecommandlockouts
\usepackage{nopageno}

\usepackage{cite}

\ifCLASSINFOpdf
\else

\fi

\usepackage{siunitx}
\usepackage{amsmath}
\usepackage{amssymb}
\usepackage{leftindex}
\usepackage{multirow}
\usepackage{arydshln}
\usepackage{comment}
\usepackage{bm}
\usepackage{booktabs}
\usepackage{graphicx}

\usepackage{pgfplots}
\pgfplotsset{compat=1.18}

\usepackage{amsthm}
\theoremstyle{definition}

\theoremstyle{remark}

\def  \pdd {prograph-based code with local irregularity}
\def  \pddAbr {PLI}
\def  \pexitnew {PLI-EXIT}
\def  \vnAbr {VN}
\def  \cnAbr {CN}

\usepackage{tikz}

\usetikzlibrary{fit}
\tikzset{%
  highlight_red/.style={rectangle,rounded corners,fill=red!15,draw,fill opacity=0.5,thick,inner sep=0pt}
}

\tikzset{%
  highlight_blue/.style={rectangle,rounded corners,fill=blue!15,draw,fill opacity=0.5,thick,inner sep=0pt}
}

\usetikzlibrary{plotmarks}

\usepackage{multicol}
\usepackage{lipsum}
\newcounter{MYtempeqncnt}

\begin{document}

%
\definecolor{kit-green100}{rgb}{0,.59,.51}
\definecolor{kit-green70}{rgb}{.3,.71,.65}
\definecolor{kit-green50}{rgb}{.50,.79,.75}
\definecolor{kit-green30}{rgb}{.69,.87,.85}
\definecolor{kit-green15}{rgb}{.85,.93,.93}
\definecolor{KITgreen}{rgb}{0,.59,.51}

\definecolor{KITpalegreen}{RGB}{130,190,60}
\colorlet{kit-maigreen100}{KITpalegreen}
\colorlet{kit-maigreen70}{KITpalegreen!70}
\colorlet{kit-maigreen50}{KITpalegreen!50}
\colorlet{kit-maigreen30}{KITpalegreen!30}
\colorlet{kit-maigreen15}{KITpalegreen!15}

\definecolor{KITblue}{rgb}{.27,.39,.66}
\definecolor{kit-blue100}{rgb}{.27,.39,.67}
\definecolor{kit-blue70}{rgb}{.49,.57,.76}
\definecolor{kit-blue50}{rgb}{.64,.69,.83}
\definecolor{kit-blue30}{rgb}{.78,.82,.9}
\definecolor{kit-blue15}{rgb}{.89,.91,.95}

\definecolor{KITyellow}{rgb}{.98,.89,0}
\definecolor{kit-yellow100}{cmyk}{0,.05,1,0}
\definecolor{kit-yellow70}{cmyk}{0,.035,.7,0}
\definecolor{kit-yellow50}{cmyk}{0,.025,.5,0}
\definecolor{kit-yellow30}{cmyk}{0,.015,.3,0}
\definecolor{kit-yellow15}{cmyk}{0,.0075,.15,0}

\definecolor{KITorange}{rgb}{.87,.60,.10}
\definecolor{kit-orange100}{cmyk}{0,.45,1,0}
\definecolor{kit-orange70}{cmyk}{0,.315,.7,0}
\definecolor{kit-orange50}{cmyk}{0,.225,.5,0}
\definecolor{kit-orange30}{cmyk}{0,.135,.3,0}
\definecolor{kit-orange15}{cmyk}{0,.0675,.15,0}

\definecolor{KITred}{rgb}{.63,.13,.13}
\definecolor{kit-red100}{cmyk}{.25,1,1,0}
\definecolor{kit-red70}{cmyk}{.175,.7,.7,0}
\definecolor{kit-red50}{cmyk}{.125,.5,.5,0}
\definecolor{kit-red30}{cmyk}{.075,.3,.3,0}
\definecolor{kit-red15}{cmyk}{.0375,.15,.15,0}

\definecolor{KITpurple}{RGB}{160,0,120}
\colorlet{kit-purple100}{KITpurple}
\colorlet{kit-purple70}{KITpurple!70}
\colorlet{kit-purple50}{KITpurple!50}
\colorlet{kit-purple30}{KITpurple!30}
\colorlet{kit-purple15}{KITpurple!15}

\definecolor{KITcyanblue}{RGB}{80,170,230}
\colorlet{kit-cyanblue100}{KITcyanblue}
\colorlet{kit-cyanblue70}{KITcyanblue!70}
\colorlet{kit-cyanblue50}{KITcyanblue!50}
\colorlet{kit-cyanblue30}{KITcyanblue!30}
\colorlet{kit-cyanblue15}{KITcyanblue!15}

\title{Protograph-Based LDPC Codes\\ with Local Irregularity\\
\thanks{Funded by the European Union (Grant Agreement No.~101120422). Views and opinions expressed are, however, those of the author(s) only and do not necessarily reflect those of the European Union or REA. Neither the European Union nor the granting authority can be held responsible for them.}}

\author{\IEEEauthorblockN{Vincent Wüst, Erdem Eray Cil, and~Laurent~Schmalen}
\IEEEauthorblockA{Communications Engineering Lab, Karlsruhe Institute of Technology (KIT), 76131 Karlsruhe, Germany\\
\texttt{vincent.wuest@kit.edu}}

}

\maketitle
\thispagestyle{empty}

\begin{abstract}
Forward error correcting (FEC) codes are used in many communication standards with a wide range of requirements. FEC codes should work close to capacity, achieve low error floors, and have low decoding complexity. In this paper, we propose a novel category of low-density parity-check (LDPC) codes, based on protograph codes with local irregularity. This new code family generalizes conventional protograph-based LDPC codes and is capable of reducing the iterative decoding threshold of the conventional counterpart. We introduce an adapted version of the protograph extrinsic information transfer (PEXIT) algorithm to estimate decoding thresholds on the binary-input additive white Gaussian noise channel, perform optimizations on the local irregularity, and simulate the performance of some constructed codes.

\end{abstract}


\IEEEpeerreviewmaketitle

\section{Introduction}

\IEEEPARstart{O}{ne} of the most heavily used families of forward error correcting (FEC) codes over the last decades is the family of low-density parity-check (LDPC) codes. LDPC codes gained interest because they can perform at rates close to the capacity~\cite{irregular_ldpc} and have lower error floors than, e.g., Turbo codes~\cite{ProtSurvey}.
Being an extensively studied subject, different types of LDPC codes were developed, enforcing some structure on LDPC codes that provide enough degrees of freedom to reach favorable iterative decoding thresholds while maintaining a feasible encoding and decoding complexity, low error floors, and linear minimum Hamming distance growth.

One promising and widely used code class are protograph LDPC codes~\cite{pexit_firstpaper}. 
Multiple heuristic measures to optimize their decoding thresholds, such as introducing a precoding structure, are known to simplify the code construction~\cite{ProtSurvey}.
A useful tool to derive the iterative decoding threshold and further facilitate the code design is the protograph extrinsic information transfer (PEXIT) analysis~\cite{PEXITderivation}.
One protograph code ensemble using a precoding structure with a low decoding threshold is the accumulate-repeat-by-4-jagged-accumulate (AR4JA) code~\cite{ProtSurvey}. Besides a low decoding threshold, it also has low encoding and decoding complexity and fulfills the necessary condition for protographs to have linear minimum Hamming distance growth. However, the rate-$1/2$ AR4JA code still has a $\SI{0.4}{dB}$ gap to capacity for the binary-input additive white Gaussian noise (BI-AWGN) channel \cite[Fig.~17]{ar4japaper}.

A code family generalizing protograph-based LDPC codes are multi-edge-type LDPC (MET-LDPC) codes~\cite{metIntro}. MET-LDPC codes allow various edge types that can have different number of nodes involved, and other than for protographs, they do not have to inherently consist of regular subgraphs. Being a more general code family, they are expected to achieve better performance than protograph-based codes. However, the optimization of MET-LDPC codes must consider a large number of parameters~\cite{metIntro},~\cite{METoptimisation}, which might complicate the optimization in practice. Protographs can be seen as a subcategory of MET-LDPC codes, and hence we seek a generalization of protograph-based LDPC codes that enables a simpler optimization than MET-LDPC codes.

The generalization of protographs is based on the idea that introducing local irregularities in subgraphs might improve its decoding threshold. Local irregularities are also present in spatially coupled LDPC (SC-LDPC) codes and contribute to their outstanding performance \cite{scldpc_intro}. 
Besides the local irregularities at the boundaries, irregular SC-LDPC codes can also be designed~\cite{scldpc_irreg}. 
Another approach to introduce irregularities is to mask an existing base graph~\cite{maskingQC}. The mask determines whether the corresponding connections of a subgraph will be used in the final code.
Both SC-LDPC codes and the masking technique provide local irregularities before lifting, which can be modeled by a conventional protograph.

In this paper, we introduce a new generalization of a protograph-based LDPC code called \pdd\ (\pddAbr). \pddAbr s generalize conventional protographs by introducing a local irregularity within each edge type. Note that they are still a subclass of MET-LDPC codes. Their iterative decoding threshold can be calculated by modifying and extending the conventional PEXIT algorithm. We call the adapted version of the PEXIT algorithm \pexitnew. Using the \pexitnew\ algorithm, we optimize \pddAbr s with respect to the iterative decoding threshold. Since conventional protographs are special cases of \pddAbr s, we use the rate-$1/2$ AR4JA protograph as an initial point for optimization. By comparing simulation results of both conventional protographs and \pddAbr s over the BI-AWGN channel, we investigate the performance improvements achieved.

\vspace{-0.1cm}

\section{Protograph LDPC Codes}\label{sec:protographs}

We denote the protographs known in the literature as conventional protographs.
We first revisit conventional protographs before introducing \pddAbr s.

\subsection{Conventional Protographs}\label{subsec:conv_prot}
\begin{figure}
\centering
\begin{tikzpicture}[scale=1.2, every node/.style={circle, draw, minimum size=1cm}]
    \node (v1) at (0,2) {$\mathsf{v}_1$};
    \node (v2) at (0,1) {$\mathsf{v}_2$};
    \node (v3) at (0,0) {$\mathsf{v}_3$};

    \node[rectangle] (c1) at (2,1.5) {$\mathsf{c}_1$};
    \node[rectangle] (c2) at (2,0.5) {$\mathsf{c}_2$};

    \draw (v1.10) edge[auto=right,-] (c1.170);
    \draw (v1.0) edge[auto=right,-] (c1.180);
    \draw (v1.-10) edge[auto=right,-] (c1.190);
    \draw (v2) edge[auto=right,-] (c1);
    \draw (v2) edge[auto=right,-] (c2);
    \draw (v3.-5) edge[auto=right,-] (c2.185);
    \draw (v3.5) edge[auto=right,-] (c2.175);
    \draw (v3.20) edge[auto=right,-] (c1.210);
\end{tikzpicture}
\caption{Example of a protograph.}
\label{fig:Protograph_tannergraph}
\vspace{-0.7cm}
\end{figure}
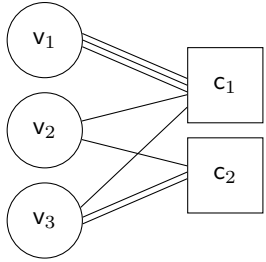

A protograph can be characterized by a graph with potentially parallel edges, with an example depicted in Fig.~\ref{fig:Protograph_tannergraph}. 
The protograph consists of variable nodes (\vnAbr s), indicated by circles, and check nodes (\cnAbr s), indicated by square boxes, that form a bipartite graph, i.e., \vnAbr s can only be connected to \cnAbr s and vice versa. We denote the total number of \cnAbr s by $m$ ($m=2$ in Fig.~\ref{fig:Protograph_tannergraph}), and the total number of \vnAbr s by $n$ ($n=3$ in Fig.~\ref{fig:Protograph_tannergraph}). An edge of the protograph can be characterized by its endpoints and multiplicity (the number of parallel edges). For example, in Fig.~\ref{fig:Protograph_tannergraph}, the edge between \vnAbr\ $\mathsf{v}_3$ and \cnAbr\ $\mathsf{c}_2$ has endpoints $(2,3)$ and multiplicity $2$. 

Equivalently, another way of characterizing a protograph is to use a protomatrix (also known as a base matrix) $\bm{B}$, with $\bm{B} \in \mathbb{N}_{0}^{m\times n}$. The entry in row $i$ and column $j$ of the protomatrix indicates the number of edges between \cnAbr\ $\mathsf{c}_i$ and \vnAbr\ $\mathsf{v}_j$, i.e., its multiplicity.
The protograph of Fig.~\ref{fig:Protograph_tannergraph} has the protomatrix
\begin{align}\label{eq:ex_prot}
\bm{B} = \begin{pmatrix}
3 & 1 & 1 \\
0 & 1 & 2 \\
\end{pmatrix}    
.\end{align}
The general set of double indices to describe the entries of $\bm{B}$ is given by
\[
\mathcal{M} := \{1,\ldots,m\}\times \{1,\ldots,n\},
\]
and the design rate ${R}$ of a protograph is given by 
\begin{equation*}
    {R} = \frac{n-m}{n-n_{\text{punc}}},
\end{equation*}
where $n_{\text{punc}}$ denotes the number of punctured bits, if any exist.
Another useful tool to analyze LDPC codes, is the concept of \vnAbr\ degrees ($d^{\text{V}}$) and \cnAbr\ degrees ($d^{\text{C}}$). The degree of a node is the number of edges incident to a node. For example, in Fig.~\ref{fig:Protograph_tannergraph}, \cnAbr\ $\mathsf{c}_1$ has check node degree $5$ when counting each edge at its multiplicity. \vnAbr\ $\mathsf{v}_1$ has variable node degree $3$ etc. 

To obtain a binary code from a protograph, the protograph is lifted. The lifting procedure (also known as the copy-permute procedure) yields a parity-check matrix (PCM) $\bm{H}$ from the base matrix $\bm{B}$. Graphically, the lifting procedure can be thought of as copying the base graph $S$ times and then permuting edges that originate from the same  \vnAbr\ to \cnAbr\ connection $(i,j)$ in the base graph such that parallel edges are eliminated.
In matrix form, we can lift~\eqref{eq:ex_prot} by replacing each entry $B_{i,j}$ (the value of $\bm{B}$ at position $(i,j)$) with the sum of $B_{i,j}$  randomly chosen, non-overlapping permutation matrices of size $S\times S$. The final code $\bm{H}$ is of dimension $Sm\times Sn$ and consists of $mn$ block matrices of dimension $S\times S$. We denote the $S\times S$ subgraph of $\bm{H}$, derived from the entry $(i,j)$ in $\bm{B}$, as the local rate-$0$ code $\bm{H}_{i,j}$.

\subsection{Iterative Decoding Threshold}\label{subsec:it_dec_thr}
One of the most important performance indicators for protographs is the iterative decoding threshold~\cite{ProtSurvey}. A common tool to determine the iterative decoding threshold is the PEXIT algorithm~\cite{PEXITderivation}. The BI-AWGN channel is completely characterized by the fraction $E_{\text{b}}/N_0$, i.e., the energy needed per information bit $E_{\text{b}}$ divided by the channel noise power density $N_{0}$. The PEXIT algorithm determines whether an iterative decoder converges for a given value of $E_{\text{b}}/N_0$. 
The update equations to determine the decoding threshold are given in~\eqref{eq:conventional_muCH}-\eqref{eq:conventional_IApp}. We quickly summarize the PEXIT algorithm~\cite{PEXITderivation}. First, we introduce the $J$-function needed to map the mean of the Gaussian messages to the extrinsic mutual information (MI)~\cite{Jfunction} as
\begin{equation}
    J(\mu) := 1 - \int_{-\infty}^{\infty}\frac{\text{e}^{-(\tau-\mu)^{2}/ (4\mu)}}{\sqrt{4\pi\mu}}\log_{2}(1+\text{e}^{-\tau})\text{d}\tau
.\end{equation}

\begin{figure*}[!t]

\normalsize

\begin{align}\label{eq:conventional_muCH}
I_{\text{Ch}}(j) &= J\left(4{R}\frac{E_\text{b}}{N_0}\right) , \quad \forall j \in \{1,\ldots,n\} \text{ and $j$ not punctured} \text{, else } I_{\text{Ch}}(j) = 0 .
\\\label{eq:conventional_IEV}
    I_{(\ell)}^{\mathrm{EV}}(i,j) &= J\left(\sum_{s=1}^{m}b_{s,j}J^{-1}\left(I_{(\ell - 1)}^{\mathrm{EC}}(s,j)\right)-J^{-1}\left(I_{(\ell-1)}^{\mathrm{EC}}(i,j)\right)+J^{-1}\left(I_{\text{Ch}}(j)\right)\right), \quad \forall (i,j) \in \mathcal{M}
\\\label{eq:conventional_IEC}
    I_{(\ell)}^{\mathrm{EC}}(i,j) &= 1 - J\left(\sum_{t=1}^{n}b_{i,t}J^{-1}\left(1-I_{(\ell)}^{\mathrm{EV}}(i,t)\right) -J^{-1}\left(1-I_{(\ell)}^{\mathrm{EV}}(i,j)\right)\right), \quad \forall (i,j) \in \mathcal{M}
\\\label{eq:conventional_IApp}
    I_{(\ell)}^{\text{App}}(j) &= J\left(J^{-1}\bigl(I_{\text{Ch}}(j)\bigr)+\sum_{s=1}^{m}b_{s,j}J^{-1}\left({I_{(\ell)}^{\mathrm{EC}}(s,j)}\right)\right) , \quad \forall j \in \{1,\ldots,n\}
\end{align}

\hrulefill

\vspace*{-0.65cm}
\end{figure*}

Now we have all the tools to introduce the conventional PEXIT algorithm:
\begin{enumerate}
    \item \textbf{Initialization step:} The channel MI is given in~\eqref{eq:conventional_muCH}. \vnAbr s that are punctured do not carry any information. The messages between \vnAbr s and \cnAbr s are initialized to zero for the first iteration $\ell = 0$.
    \item \textbf{Variable to check update $I^{\mathrm{EV}}$:} The MI of the messages that are passed from \vnAbr s to \cnAbr s is given by~\eqref{eq:conventional_IEV} for iterations with $\ell \geq 1$. Only edges with base matrix entry different from zero are considered.
    \item \textbf{Check to variable update $I^{\mathrm{EC}}$:} The MI of the messages that are passed from \cnAbr s to \vnAbr s are given by~\eqref{eq:conventional_IEC} for iterations with $\ell \geq 1$. Only edges with base matrix entry different from zero are considered.
    \item \textbf{A-posteriori MI $I^{\mathrm{App}}$:} The a-posteriori MI evaluates to~(\ref{eq:conventional_IApp}) for iterations with $\ell \geq 1$. If the a-posteriori MI is sufficiently close to $1$ for all \vnAbr s, the algorithm stops. Otherwise, we repeat Step 2).
\end{enumerate}
For a given protograph, PEXIT is said to converge if $I^{\text{App}}$ converges to $1$ for all variable nodes $\mathsf{v}_j$ within some predefined number of iterations.

\subsection{Protographs with Local Irregularity}\label{subsec:PLDD}

First, we extend the concept of node degrees to local node degrees. The number of edges between a \cnAbr\ $\mathsf{c}_i$ and a specific \vnAbr\ $\mathsf{v}_j$, can be denoted as the local \vnAbr\ degree $d_{i,j}^{\text{V}}$ or the local \cnAbr\ degree $d_{i,j}^{\text{C}}$ respectively. Local degrees can also be stated in the following way: The local node degree of the edge $(i,j)$ corresponds to the total node degree of the same node considering only the subgraph associated to $\bm{H}_{i,j}$. For a conventional protograph, the local check node degree $d_{i,j}^{\text{C}}$ as well as the local check node degree $d_{i,j}^{\text{V}}$ between \cnAbr\ $\mathsf{c}_i$ and \vnAbr\ $\mathsf{v}_j$ are equivalent to the multiplicity of the edge $(i,j)$ in~$\bm{B}$.

The main idea of \pddAbr s is to take a conventional protograph and then introduce some additional degrees of freedom, by allowing the local \vnAbr\ degrees to be irregular, while keeping the local \cnAbr\ degrees fixed to the values of the base graph. We do so in the following way:
Similarly to the conventional case, we make use of a base matrix $\bm{B}$ from which we can directly deduce the number of variable nodes and check nodes of the protograph. However, this time the base matrix entry at position $(i,j)$ does not simply represent the edge multiplicity between \cnAbr\ $\mathsf{c}_i$ and \vnAbr\ $\mathsf{v}_j$, but it represents the local non-negative integer check node degree $d_{i,j}^{\text{C}}$. Every base matrix of a conventional protograph can be the protograph of a \pddAbr.
Here we state a general base matrix:
\[
\bm{B} = \begin{pmatrix}
d^{\text{C}}_{1,1} & d^{\text{C}}_{1,2} & \cdots & d^{\text{C}}_{1,n} \\
d^{\text{C}}_{2,1} & d^{\text{C}}_{2,2} & \cdots & d^{\text{C}}_{2,n} \\
\vdots & \vdots & \ddots & \vdots \\
d^{\text{C}}_{m,1} & d^{\text{C}}_{m,2} & \cdots & d^{\text{C}}_{m,n}
\end{pmatrix}.
\]

\begin{figure*}[!t]

\normalsize

\setcounter{MYtempeqncnt}{\value{equation}}

\setcounter{equation}{12}
\begin{align}
\label{eq:I_EV}
     I_{(\ell)}^{\mathrm{EV}}(i,j) &= \sum_{\mathclap{\substack{k_{i,j}\in\Omega\\ k_{u,v} \in \Omega \text{, } \forall (u,v)\in \mathcal{M}_{\neq i,j}}}}\lambda_{i,j}(k_{i,j})L_{\mathcal{M}_{\neq i,j}}(\mathcal{K}_{\neq i,j})\left[J \left( J^{-1}\bigl(I_{\text{Ch}}(j)\bigr) \Bigr. 
     \Bigl. -J^{-1}\left(I_{(\ell-1)}^{\mathrm{EC}}(i,j)\right)+\sum_{s=1}^{m}k_{s,j}J^{-1}\left(I_{(\ell-1)}^{\mathrm{EC}}(s,j)\right)\right)\right]\\
    \mathcal{M}_{\neq i,j} :&= \{(u,v) \in \{1,\ldots,m\}\times\{1,\ldots,n\}| u\neq i, v=j \} \text{, } \mathcal{K}_{\neq i,j} := \{k_{u,v}\in \Omega|  (u,v)\in \mathcal{M}_{\neq i,j} \}, \quad \forall (i,j) \in \mathcal{M} \nonumber
\\\label{eq:I_EC}
    I_{(\ell)}^{\mathrm{EC}}(i,j) &= 1 - J\left(\sum_{t=1}^{n}d^{\text{C}}_{i,t}J^{-1}\left(1-I_{(\ell)}^{\mathrm{EV}}(i,t)\right) -J^{-1}\left(1-{I_{(\ell)}^{\mathrm{EV}}(i,j)}\right)\right),\\ \text{ if } d^{\text{C}}_{i,j} &\neq 0 \text{, else  }I_{(\ell)}^{\mathrm{EC}}(i,j) = 0, \quad \forall (i,j) \in \mathcal{M}\nonumber
\\\label{eq:I_app}
    I_{(\ell)}^{\text{App}}(j) &= 
    \sum_{\mathclap{k_{u,v}\in\Omega\text{, } \forall (u,v)\in \mathcal{M}_{\cdot,j}}} L_{\mathcal{M}_{\cdot,j}}(\mathcal{K}_{\cdot,j})\left[J\left(J^{-1}\bigl(I_{\text{Ch}}(j)\bigr)+\sum_{s=1}^{m}k_{s,j}J^{-1}\left(I_{(\ell)}^{\mathrm{EC}}(s,j)\right)\right)\right]\\
    \mathcal{M}_{\cdot,j} :&= \{(u,v) \in \{1,\ldots,m\}\times\{1,\ldots,n\}| v = j \} \text{, } \mathcal{K}_{\cdot,j} := \{k_{u,v}\in \Omega|  (u,v)\in \mathcal{M}_{\cdot,j} \}, \quad \forall j \in \{1,\ldots,n\} \nonumber
\end{align}

\setcounter{equation}{\value{MYtempeqncnt}}

\hrulefill

\vspace*{-0.65cm}
\end{figure*}

On the other side, we allow different \vnAbr s within one lifted subgraph $\bm{H}_{i,j}$ to take on different local \vnAbr\ degrees $d^{\text{V}}_{i,j}$. The frequency of occurrence of a certain local degree within $\bm{H}_{i,j}$ can be described by the local \vnAbr\ degree distribution $L_{i,j}(\cdot)$. Typically, we restrict the local \vnAbr\ degrees to take on values within the set $\Omega = \{1,\ldots,d_{\text{max}}\}$. The local \vnAbr\ degree distribution is a mapping 
\begin{equation}\label{eq:partial_dd_def}
    L_{i,j}: \Omega \rightarrow [0,1] , \quad \forall (i,j) \in \mathcal{M} .
\end{equation} 
We can think of $L_{i,j}(k)$ as the fraction of \vnAbr s in $\bm{H}_{i,j}$ of degree $k$. 
The local \vnAbr\ degree distribution must fulfill

\begin{equation}\label{eq:L_sum_one}
    \sum_{k \in \Omega} L_{i,j}(k) = 1 , \quad \forall (i,j) \in \mathcal{M}.
\end{equation}

The number of outgoing edges from a \vnAbr\ must match the edges that are connected to a \cnAbr\ in the lifted graph, leading to the condition
\begin{equation}\label{eq:dd_constraint}
\sum_{k=1}^{d_{\text{max}}}kL_{i,j}(k) = d^{\text{C}}_{i,j} , \quad \forall (i,j) \in \mathcal{M}    
.\end{equation}

 Furthermore, to construct the \pexitnew\ algorithm, we also need to consider the edge perspective of the degree distributions. The frequency of occurrence that an edge of subgraph $\bm{H}_{i,j}$ will be connected to a variable node of degree $k$ in $\bm{H}_{i,j}$ can be described by the local edge perspective degree distribution
 \begin{equation}\label{eq:lambda}
    \lambda_{i,j}: \Omega \rightarrow [0,1] , \quad \forall (i,j) \in \mathcal{M} .
\end{equation}

The following relationship exists between variable node and edge perspective local degree distributions:
\begin{align}\label{eq:lambda_dep}
    \quad \lambda_{i,j}(k) &\stackrel{(a)}{=} \frac{SkL_{i,j}(k)}{S\sum_{\hat{k}\in \Omega}\hat{k}L_{i,j}(\hat{k})} \stackrel{(b)}{=} \frac{kL_{i,j}(k)}{d^{\text{C}}_{i,j}},\\  
    \forall k &\in \Omega, \quad \forall (i,j) \in \mathcal{M} \nonumber .
\end{align}

The first equality $(a)$ follows from counting all the edges of subgraph $\bm{H}_{i,j}$. Then, $(b)$ follows from~\eqref{eq:dd_constraint}. It can be shown that
\begin{equation}\label{eq:lambda_sum_to_one}
\sum_{k \in \Omega} \lambda_{i,j}(k) = \sum_{k\in \Omega}\frac{kL_{i,j}(k)}{d^{\text{C}}_{i,j}} =1, \quad \forall (i,j) \in \mathcal{M} .
\end{equation}
So far, we only considered local degree distributions of one subgraph $\bm{H}_{i,j}$ at a time. If we want to make combined statements about a local \vnAbr\ degree $d^{\text{V}}_{i,j}$ in $\bm{H}_{i,j}$ and a local \vnAbr\ degree $d^{\text{V}}_{u,j}$ in $\bm{H}_{u,j}$ for some distinct edges $(i,j)$ and $(u,j)$, we can consider the joint local degree distribution defined on the set $\Omega \times \Omega$. In the protograph, this joint local degree distribution would correspond to the degree distribution of \vnAbr\ $\mathsf{v}_j$ arising from the connections to \cnAbr s $\mathsf{c}_i$ and $\mathsf{c}_u$. The joint local degree distribution allows us to express the frequency of occurrence of any particular realization of local \vnAbr\ degrees $d^{\text{V}}_{i,j}$ and $d^{\text{V}}_{u,j}$. In the construction of the PCM $\bm{H}$, we lift the subgraphs $\bm{H}_{i,j}$ and $\bm{H}_{u,j}$ independently and model the joint local degree distribution as the multiplication of the individual local degree distributions. The joint local degree distributions still map to the interval $[0,1]$.

For example, say $d^{\text{V}}_{i,j}$ takes values in $\{2, 4\}$ with \mbox{$L_{i,j}(2) = 1 - L_{i,j}(4) = 1/2$} and $d^{\text{V}}_{u,j}$ takes values in $\{1,7\}$ with $L_{u,j}(1) = 1-L_{u,j}(7) = 1/6$. Then, when picking one local \vnAbr\ degree from subgraph $\bm{H}_{i,j}$ and one from subgraph $\bm{H}_{u,j}$, the selected pair $(d^{\text{V}}_{i,j}$, $d^{\text{V}}_{u,j})$ will be equal to $(2,1)$ in $1$ out of $12$ times, and it will be equal to $(2,7)$ in $5$ out of $12$ times etc.

By describing different combinations of local \vnAbr\ degrees from different subgraphs and by describing the frequency of their occurrence, we can model the core difference from conventional protographs to \pddAbr s in the extended version of the PEXIT algorithm. While conventional protographs consist of regular subgraphs, where only one combination of local \vnAbr\ degrees from various subgraphs exist, those combinations can vary in the setting of \pddAbr s and the MI of messages being passed must be weighted according the joint local \vnAbr\ degree distribution.

Hence, we extend the concept of joint local \vnAbr\ degree distributions to more than two distinct base matrix indices. For that purpose, let us consider a general set of double indices that must fulfill a restrictive condition $\mathcal{R}$, which will enable us to consider a certain subset of protomatrix indices, e.g., all the \cnAbr s connected to \vnAbr\ $\mathsf{v}_j$ except \cnAbr\ $\mathsf{c}_i$. We can formally define
\[
\mathcal{M}_{\mathcal{R}} := \{(u,v) \in \{1,\ldots,m\}\times\{1,\ldots,n\}:(u,v)\hspace{0.25em}\text{fulfills}\hspace{0.25em} \mathcal{R}\} ,
\]
and a realization of local degrees from each double index of the set of interest
\[
\mathcal{K}_{\mathcal{R}} := \{k_{u,v}\in \Omega:  (u,v)\in \mathcal{M}_{\mathcal{R}} \} ,
\]
such that
\[
L_{\mathcal{M}_{\mathcal{R}}}(\mathcal{K}_{\mathcal{R}}) := \prod_{(i,j) \in \mathcal{M}_{\mathcal{R}}}L_{i,j}(k_{i,j}) \quad \text{, with } k_{i,j} \in \Omega ,
\]
for all $(i,j) \in \mathcal{M}_{\mathcal{R}}$.

The difference to conventional protographs during code construction manifests itself in the lifting step, when the edges are rearranged to remove parallel edges. The nodes and edges must be matched in a way that within each subgraph, the frequency at which \vnAbr s of a certain degree occur matches the \vnAbr\ perspective local degree distribution.

\subsection{\pexitnew}\label{sec:pexit}
A few adaptions are necessary to generalize the PEXIT algorithm to the \pexitnew\ algorithm. The \pexitnew\ equations are given in \eqref{eq:I_EV}-\eqref{eq:I_app}.
The initialization step and \eqref{eq:I_EC} are identical to the conventional case. For the \vnAbr\ to \cnAbr\ update, the MI of the messages that must be considered depends on the various possible combinations of local \vnAbr\ degrees, which must be weighted according to the joint local \vnAbr\ degree distribution.
The MI of messages from \vnAbr\ $\mathsf{v}_j$ to \cnAbr\ $\mathsf{c}_i$ can be calculated by averaging over the local \vnAbr\ degree distributions of all the MI of messages from the incoming edges from \cnAbr s different from $\mathsf{c}_i$ (i.e., $L_{\mathcal{M}_{\neq i,j}}(\mathcal{K}_{\neq i,j})$), and by averaging over the local edge perspective degree distributions of the MI of the outgoing message from \vnAbr\ $\mathsf{v}_j$ to \cnAbr\ $\mathsf{c}_i$ (i.e., $\lambda_{i,j}(k_{i,j})$). For the latter, we consider the edge perspective, because we want to determine the MI of a message on a single edge (despite that the multiplicity might be larger).
For the \cnAbr\ to \vnAbr\ update, no averaging mechanism is required because the check nodes are regular, i.e., the amount of incoming edges and outgoing edges are fixed as they are for conventional protographs.
The a-posteriori MI needs to be updated in a fashion similar to the \vnAbr\ to \cnAbr\ update. However, as there is only MI of incoming messages to be considered, we average over all the local \vnAbr\ degree distributions (i.e., $L_{\mathcal{M}_{\cdot,j}}(\mathcal{K}_{\cdot,j})$).

\vspace{-0.1cm}
\section{Local Degree Optimization}\label{sec:opt}
\setcounter{equation}{15}
In this paper, we do not optimize any conventional protographs. We assume that an optimized conventional protograph is given. Then, we optimize the local degree distributions of the local \vnAbr\ degrees of a given base matrix.
The conventional protograph used for optimization is the rate-$1/2$ AR4JA protograph~\cite{ar4japaper} with protomatrix

\begin{equation}\label{eq:base_m_ar4ja}
\bm{B} = \begin{pmatrix}
1 & 2 & 0 & 0 & 0\\
0 & 3 & 1 & 1 & 1\\
0 & 1 & 2 & 2 & 1 \\
\end{pmatrix},
\end{equation}
where the second variable node is punctured.

Given a protograph, we optimize local degree distributions on one or multiple edges of the protograph.
The dimensionality of the optimization problem increases rapidly with the number of local degree distributions. Therefore, we will first consider the optimization of a single local degree distribution and later extend to the optimization of multiple edges. To perform the optimization introduced in the following, we always carry out $\num{500}$ iterations in the \pexitnew\ algorithm.

\subsection{Single-Edge Degree Distribution Optimization}\label{subsec:se_opt}
The cost function in the optimization problem is the decoding threshold, evaluated by \pexitnew. The optimization variables are elements of the local \vnAbr\ degree distribution $L(\cdot)$ of the edge that we want to optimize. The optimization is carried out with respect to the constraints~\eqref{eq:L_sum_one} and~(\ref{eq:dd_constraint}). In this paper, we use the genetic algorithm provided by MATLAB to solve the optimization problem. The local \vnAbr\ degrees $d^{\text{V}}_{i,j}$ take values in $\Omega = \{1,\ldots,20\}$. With this optimization technique, we generate the code $\mathcal{C}_1$, which is based on the base matrix~(\ref{eq:base_m_ar4ja}), with the optimized edge index being $(3,4)$. Table~\ref{tab:optimized_ldd} shows the local \vnAbr\ degree distribution of the single entry that has been optimized.

\subsection{Multi-Edge Degree Distribution Optimization}\label{subsec:me_opt}
When optimizing multiple local degree distributions at the same time, the dimensionality of the optimization problem increases. To simplify the optimization, we perform element-wise optimization. In an iterative process, we optimize one edge after the other and only stick to the optimized local degree distribution if it outperforms the corresponding regular edge. We repeat this process until we do not see any further improvement. In each element-wise optimization we use the same optimizer and the same $\Omega$ as in the single-edge case. The edges that improved the decoding threshold when applying a non-trivial local \vnAbr\ degree distribution are $(2,2)$, $(3,3)$ and $(3,4)$. This procedure leads to the code $\mathcal{C}_2$. 
The relevant local \vnAbr\ degree distributions are shown in Table~\ref{tab:optimized_ldd}.

\begin{table}[]
    \centering
    \caption{Optimized Local \vnAbr\ Degree Distributions. Only rows where at least one local degree is greater than $0.01$ are given.}
\begin{tabular}{ p{0.5cm} p{1.2cm} p{1.2cm} p{1.2cm} p{1.2cm}  }

 \toprule
 \multirow{2}{*}{$k$} & \multicolumn{1}{c}{$\mathcal{C}_1$} & \multicolumn{3}{c}{$\mathcal{C}_2$} \\ 
\cmidrule(lr){2-2}
\cmidrule(lr){3-5}
    &\vphantom{$L_{3,4}\left(\frac 1k \right)$} $L_{3,4}(k)$&$L_{2,2}(k)$ & $L_{3,3}(k)$ & $L_{3,4}(k)$\\
 \midrule 
 
 1& 0.94395  & 0.74712 &0.79931&0.07360\\
 2 & 0.00014& 0.00063& 0.00160& 0.85803\\
 3 & 0.00001& 0.00154& 0.00064& 0.06717\\
 6 &0.00002 & 0.00277&  0.19684&0.00026\\
 9 &0.00002 & 0.24029&  0.00004&0.00001\\
 19 &0.05408 & 0.00018&  0.00005&0.00003\\
 \bottomrule
\end{tabular}
    
    \label{tab:optimized_ldd}
    \vspace{-0.5cm}
\end{table}

\section{Simulation Results}\label{sec:results}
After optimizing the local degree distribution for the rate-$1/2$ AR4JA code, we generate a PCM by lifting the protograph. In this paper, we use random permutation matrices for lifting with a lifting factor of $S = \num{100000}$ which leads to a block length of $\num{400000}$. Such block lengths are used in high-througput (e.g., optical) communications and continuous-variable quantum key distribution \cite{ldpc_qkd}. After lifting, all $4$-cycles were removed, as in~\cite{remove_4_loop}. We use sum-product decoding with $\num{1000}$ decoding iterations to decode after transmission over a BI-AWGN channel.

\begin{figure}
    \centering
    \includegraphics{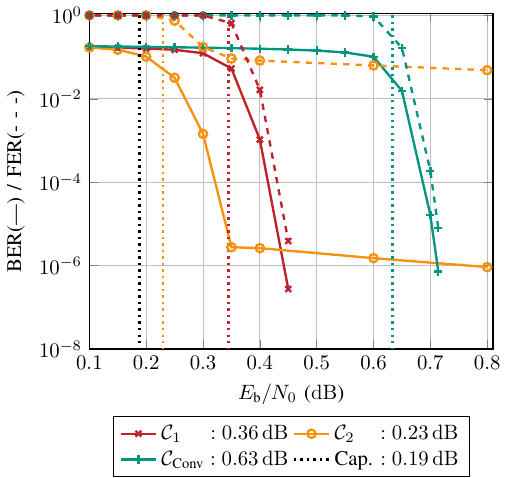}
    
    \caption{Simulated error rates as a function of $E_{\text{b}}/N_0$ for sum-product decoding. For each legend entry, the decoding threshold/ capacity is stated.}
    \label{fig:BERnFER}
\vspace{-0.5cm}
    
\end{figure}

In Fig.~\ref{fig:BERnFER}, we show the bit error and the frame error rates. The iterative decoding thresholds are shown as vertical dotted lines. The threshold of code $\mathcal{C}_2$ was roughly $\SI{0.4}{\dB}$ better than for the conventional protograph and only $\SI{0.04}{\dB}$ below capacity. In case of optimizing a single edge (code $\mathcal{C}_1$), the decoding threshold improves by roughly $\SI{0.3}{\dB}$ and is only $\SI{0.16}{\dB}$ below capacity.

The error rates of both codes $\mathcal{C}_1$ and $\mathcal{C}_2$ start to drop at their iterative decoding thresholds. However, it should be noted that the error rates of code $\mathcal{C}_2$ decay less quickly than those of the conventional protograph-based code $\mathcal{C}_{\text{Conv}}$ constructed with~\eqref{eq:base_m_ar4ja}. The code $\mathcal{C}_2$ also has a higher error floor than the other codes, for which no frame errors were observed at larger values of $E_{\text{b}}/N_{0}$ during our simulation. When investigating the dominant error patterns within the error floor, we observe that in most cases they involve \vnAbr s with degree $2$. It is known in the literature~\cite{ProtSurvey,ar4japaper,ProtRateAdaptive}, that a code with too many degree-$2$ \vnAbr s is prone to have an error floor and lacks linear minimum Hamming distance growth. The conventional rate-$1/2$ AR4JA code already contains as many degree-$2$ \vnAbr s as possible to fulfill this condition. Both $\mathcal{C}_1$ and $\mathcal{C}_2$ violate this condition, as the local \vnAbr\ degree distributions introduce additional degree-$2$ \vnAbr s.

\vspace{-0.2cm}
\section{Conclusion}\label{sec:conclusion}
In this work, we introduced protographs with local irregularity, which generalize the concept of conventional protographs by introducing local degree distributions on the edges of the base graph.
This leads to a higher degree of freedom when optimizing the code while maintaining a structure that is suitable for optimization. To do so, we use the \pexitnew\ algorithm that extends the PEXIT algorithm by incorporating the generalized model structure. The results obtained by optimizing the local degree distributions significantly outperform the conventional protograph-based code. The reference rate-$1/2$ AR4JA code was outperformed in terms of the decoding threshold by $\SI{0.3}{\dB}$ and $\SI{0.4}{\dB}$ for different codes we obtained. The error floor observed for the code with an improvement of $\SI{0.4}{\dB}$ in the decoding threshold can most likely be explained by the large number of degree-$2$ \vnAbr s.

\ifCLASSOPTIONcaptionsoff
  \newpage
\fi


\end{document}